\documentclass[aps,prl,twocolumn,groupedaddress]{revtex4}

\usepackage[pdftex]{graphicx}
\def\ccpip {CC$1\pi^+$ }
\def\ccqe {CCQE }
\def\ccpiplike {CC$1\pi^+$-like }
\def\ccqelike {CCQE-like }
\def\ccpin {CC$1\pi^0$ }

\begin{document}


\title{Measurement of the \(\nu_\mu\) charged current \(\pi^+\) to quasi-elastic cross section ratio on mineral oil in a 0.8 GeV neutrino beam}



\author{\vspace{-2mm}A.~A. Aguilar-Arevalo$^{5,*}$, C.~E.~Anderson$^{19}$,
	A.~O.~Bazarko$^{16}$, S.~J.~Brice$^{7}$, B.~C.~Brown$^{7}$,
        L.~Bugel$^{5}$, J.~Cao$^{15}$, L.~Coney$^{5}$,
        J.~M.~Conrad$^{5,13}$, D.~C.~Cox$^{10}$, A.~Curioni$^{19}$,
        Z.~Djurcic$^{5}$, D.~A.~Finley$^{7}$, B.~T.~Fleming$^{19}$,
        R.~Ford$^{7}$, F.~G.~Garcia$^{7}$,
        G.~T.~Garvey$^{11}$, C.~Green$^{7,11}$, J.~A.~Green$^{10,11}$,
        T.~L.~Hart$^{4}$, E.~Hawker$^{3,11}$,
        R.~Imlay$^{12}$, R.~A. ~Johnson$^{3}$, G.~Karagiorgi$^{5,13}$,
        P.~Kasper$^{7}$, T.~Katori$^{10,13}$, T.~Kobilarcik$^{7}$,
        I.~Kourbanis$^{7}$, S.~Koutsoliotas$^{2}$, E.~M.~Laird$^{16}$,
        S.~K.~Linden$^{19}$,J.~M.~Link$^{18}$, Y.~Liu$^{15}$,
        Y.~Liu$^{1}$, W.~C.~Louis$^{11}$,
        K.~B.~M.~Mahn$^{5}$, W.~Marsh$^{7}$, 
	V.~T.~McGary$^{5,13}$, G.~McGregor$^{11}$,
        W.~Metcalf$^{12}$, P.~D.~Meyers$^{16}$,
        F.~Mills$^{7}$, G.~B.~Mills$^{11}$,
        J.~Monroe$^{5}$, C.~D.~Moore$^{7}$, R.~H.~Nelson$^{4}$,
	P.~Nienaber$^{17}$, J.~A.~Nowak$^{12}$,
	B.~Osmanov$^{8}$, S.~Ouedraogo$^{12}$, R.~B.~Patterson$^{16}$,
        D.~Perevalov$^{1}$, C.~C.~Polly$^{9,10}$, E.~Prebys$^{7}$,
        J.~L.~Raaf$^{3}$, H.~Ray$^{8,11}$, B.~P.~Roe$^{15}$,
	A.~D.~Russell$^{7}$, V.~Sandberg$^{11}$, R.~Schirato$^{11}$,
        D.~Schmitz$^{5}$, M.~H.~Shaevitz$^{5}$, F.~C.~Shoemaker$^{16,\dagger}$,
        D.~Smith$^{6}$, M.~Soderberg$^{19}$,
        M.~Sorel$^{5,\ddagger}$,
        P.~Spentzouris$^{7}$, J.~Spitz$^{19}$, I.~Stancu$^{1}$,
        R.~J.~Stefanski$^{7}$, M.~Sung$^{12}$, H.~A.~Tanaka$^{16}$,
        R.~Tayloe$^{10}$, M.~Tzanov$^{4}$,
        R.~Van~de~Water$^{11}$, 
	M.~O.~Wascko$^{12,\S}$,
	 D.~H.~White$^{11}$,
        M.~J.~Wilking$^{4}$, H.~J.~Yang$^{15}$,
        G.~P.~Zeller$^{5,11}$, E.~D.~Zimmerman$^{4}$ \\
\smallskip
(The MiniBooNE Collaboration)
\smallskip
}
\smallskip
\smallskip
\affiliation{
$^1$University of Alabama; Tuscaloosa, AL 35487 \\
$^2$Bucknell University; Lewisburg, PA 17837 \\
$^3$University of Cincinnati; Cincinnati, OH 45221\\
$^4$University of Colorado; Boulder, CO 80309 \\
$^5$Columbia University; New York, NY 10027 \\
$^6$Embry Riddle Aeronautical University; Prescott, AZ 86301 \\
$^7$Fermi National Accelerator Laboratory; Batavia, IL 60510 \\
$^8$University of Florida; Gainesville, FL 32611 \\
$^9$University of Illinois; Urbana, IL 61801 \\
$^{10}$Indiana University; Bloomington, IN 47405 \\
$^{11}$Los Alamos National Laboratory; Los Alamos, NM 87545 \\
$^{12}$Louisiana State University; Baton Rouge, LA 70803 \\
$^{13}$Massachusetts Institute of Technology; Cambridge, MA 02139 \\
$^{15}$University of Michigan; Ann Arbor, MI 48109 \\
$^{16}$Princeton University; Princeton, NJ 08544 \\
$^{17}$Saint Mary's University of Minnesota; Winona, MN 55987 \\
$^{18}$Virginia Polytechnic Institute \& State University; Blacksburg, VA
24061\\
$^{19}$Yale University; New Haven, CT 06520\\
*Present address: Inst. de Ciencias Nucleares, Univ. Nacional Aut\'onoma de M\'exico, D.F. 04510, M\'exico \\
$\dagger$Deceased \\
$\ddagger$Present address: IFIC, Universidad de Valencia and CSIC, Valencia 46071, Spain \\
$\S$Present address: Imperial College; London SW7 2AZ, United Kingdom
}\date{\today}

\begin{abstract}
\vspace{-2mm}
Using high statistics samples of charged current $\nu_\mu$ interactions, MiniBooNE reports a measurement of the single charged pion production to quasi-elastic cross section ratio on mineral oil (CH$_2$), both with and without corrections for hadron re-interactions in the target nucleus. The result is provided as a function of neutrino energy in the range 0.4 GeV $< E_\nu <$ 2.4 GeV with 11\% precision in the region of highest statistics. The results are consistent with previous measurements and the prediction from historical neutrino calculations.
\end{abstract}


\maketitle


Future neutrino oscillation experiments will operate in the 1 GeV energy range, where charged current quasi-elastic scattering (CCQE, $\nu_\mu n \rightarrow \mu^- p$) and charged current single pion production (CC$1\pi^+$, $\nu_\mu X \rightarrow \mu^- \pi^+ X'$) are the dominant interactions. Because such processes are the largest contributors to the event samples in such experiments, there has been much interest in making better determinations of their cross sections. At present, the ratio of CC$1\pi^+$/CCQE cross sections has been measured to  $\sim$30\% precision based on small event samples \cite{ANL,BNL,K2K}. A high statistics measurement of these processes necessarily requires the use of nuclear targets where final state interactions obscure the actual value of the ratio of the cross sections on nucleons. Experimentally, it is the cross section on complex nuclei including the effects of final state interactions which is more relevant (experiments can only identify particles that actually exit the struck nucleus). In this letter, MiniBooNE reports the first measurement of the observed CC$1\pi^+$/CCQE cross section ratio as a function of neutrino energy including the effects of hadronic re-interactions. Additionally, an underlying ratio at the nucleon level is extracted to facilitate comparison with prior measurements \cite{ANL,K2K}. Precise knowledge of this cross section ratio is particularly important for future $\nu_{\mu}$ disappearance searches, in which CC$1\pi^+$ events typically constitute either a class of signal events or a large background to the CCQE signal.  The uncertainty on the CC$1\pi^+$/CCQE cross section ratio therefore limits the precision of these measurements.

The Booster Neutrino Beam at Fermilab provides a neutrino source which is particularly well-suited to make this measurement; about 40\% of $\nu_\mu$ neutrino interactions in MiniBooNE are expected to be \ccqe and 24\% CC$1\pi^+$.  The beam itself is composed of 93.6\% $\nu_\mu$ with a mean energy of about 800 MeV and $5.9\%$ ($0.5\%$) $\bar{\nu}_\mu$ ($\nu_e$) contamination \cite{MB_flux}. The neutrinos are detected in the MiniBooNE detector \cite{MB_det}, a 12.2 m diameter spherical tank filled with 818 tons of undoped mineral oil located 541 m downstream of the beryllium target. At the energies relevant to this analysis, the products of the interactions produce primarily \v{C}erenkov light with a small fraction of scintillation light \cite{MB_det}. The light is detected by 1280 8-inch photomultiplier tubes (PMTs) which line the MiniBooNE inner tank.  This inner tank region is optically isolated from a surrounding veto region, instrumented with 240 PMTs, that serves to reject incoming cosmic rays and partially contained neutrino interactions.

Neutrino interactions within the detector are simulated with the v3 NUANCE event generator \cite{NUANCE}. CCQE interactions on carbon are generated using the relativistic Fermi gas model \cite{SM} tuned to better describe the observed distribution of $\nu_\mu$ CCQE interactions in MiniBooNE \cite{MB_ccqe}. Resonant CC$1\pi^+$ events are simulated using the Rein and Sehgal (R-S) model \cite{RS}, as implemented in NUANCE with an axial mass $M_A^{1\pi}=1.1$ GeV. The angular distribution of the decaying pions in the center of mass of the recoiling resonance follows the helicity amplitudes of \cite{RS}. In MiniBooNE, $87\%$ of CC$1\pi^+$ production is predicted to occur via the $\Delta(1232)$ resonance, but 17 higher mass resonances and their interferences, as well as a non-resonant background \cite{RS} that accounts for roughly $6\%$ of CC$1\pi^+$ events, are also included in the model.  Coherently produced CC$1\pi^+$ events are generated using the R-S model \cite{RS_coh} with the R-S absorptive factor replaced by NUANCE's pion absorption model and the overall cross section rescaled to reproduce MiniBooNE's recent measurement of neutral current coherent $\pi^0$ production \cite{MB_pi0}. Coherent $\pi^+$ production is predicted to compose less than 6\% of the MiniBooNE \ccpip sample due to the small coherent cross section \cite{K2K_coh,SB} and the dominance of the $\Delta^{++}$ resonance. A GEANT3-based detector model \cite{GEANT} simulates the response of the detector to particles produced in these neutrino interactions.

To select $\nu_{\mu}$ charged current interactions, simple requirements on the amount of charge detected in the tank ($> 175$ tank PMT hits) and in the veto region ($<6$ veto PMT hits), location of the event in the tank ($<500$ cm from the center of the detector), and event time (event must occur while the beam is passing through the detector) are first applied. Further requirements on the number of decay electrons in the event are then used to isolate CCQE from CC$1\pi^+$ interactions.  $\nu_\mu$ CCQE events are selected by requiring the detection of a single electron (from the decay of a stopped muon) within 100 cm of the endpoint of the muon track \cite{MB_ccqe}.  Identification of the decay electron is possible because it follows the detection of the muon by a distinct time interval. $\nu_\mu$ CC$1\pi^+$ interactions are identified by requiring the detection of two electrons (from the decay of the muon ($\mu^- \rightarrow e^- \nu_\mu \bar{\nu_e}$) and pion ($\pi^+ \rightarrow \mu^+ \nu_\mu$, $\mu^+ \rightarrow e^+ \bar{\nu_\mu} \nu_e$)), at least one of which must be within 150 cm of the endpoint of the muon track. The model dependence of the event selection is rather small since we require only that the $\mu^-$ and $\pi^+$ decay. After cuts and with $5.58 \cdot 10^{20}$ protons on target, the CCQE data sample contains 193,709 events and the CC$1\pi^+$ sample 46,172, making these the largest samples collected in this energy range by more than an order of magnitude.
      
The \ccqe and \ccpip reconstruction requires a detailed model of light production and propagation in the tank to predict the charge distribution for a given vertex and muon angle.  The muon vertex, track angle, and energy, are found with a maximal likelihood fit, with the energy being determined from the total tank charge. The neutrino energy for both samples is reconstructed from the observed muon kinematics, treating the interaction as a 2-body collision and assuming that the target nucleon is at rest inside the nucleus:
\vspace{-7mm}

\begin{equation}
\vspace{1mm}
E_\nu = \frac{1}{2}\frac{2m_pE_\mu + m_1^2 - m_p^2 -m_\mu^2}{m_p - E_\mu + \cos\theta_\mu \sqrt{E_\mu^2-m_\mu^2}}
\vspace{-2mm}
\end{equation}

\vspace{-1mm}\noindent Here \(m_p\) is the mass of the proton, \(m_\mu\) is the mass of the muon, \(m_1\) is the mass of the neutron in \ccqe events and of the \(\Delta(1232)\) in \ccpip \hspace{-1.5mm}, \(\theta_\mu\) is the reconstructed angle of the muon with respect to the beam axis (in the lab frame), and \(E_\mu\) is the reconstructed muon energy.

The distributions of signal events in neutrino energy are obtained through a two step process.  First, the aforementioned cuts are applied to select the \ccpip and \ccqe samples. These samples can be characterized by the cut efficiency (the fraction of signal events in the data set that pass the relevant cuts) and the signal fraction (the fraction of events in a given sample that are in fact signal events). Second, a Monte Carlo simulation (MC) is used to predict the signal fractions and cut efficiencies; these values are then used to correct the raw numbers of events passing cuts.

For our primary measurement, we define \ccpip signal as events with exactly one $\mu^-$ and one $\pi^+$ escaping the struck nucleus (which we call \ccpiplike events) and \ccqe signal as those with exactly one $\mu^-$ and no pions (\ccqelike events). Both event classes may include any number of protons or neutrons, but no other types of hadrons.  The observed cross section ratio is then defined as the ratio of \ccpiplike to \ccqelike events and thus has not been corrected for re-interactions in the struck nucleus.  The signal fraction of the \ccpiplike (\ccqelike) sample is predicted to be $92\%$ ($83\%$) and the cut efficiency is predicted to be $26\%$ ($38\%$) in 500 cm.  Table \ref{effective_comp} gives the composition of the \ccpiplike and \ccqelike signal events in the MC.

\begin{table}[htb]
\begin{center}
\vspace{-1mm}
\begin{tabular}{| l | c | c |}
	\hline
	& Fraction of $CC1\pi^+$- \vspace{-1.1mm}& Fraction of CCQE- \\
	Process & like events (\%) & like events (\%)\\ \hline \hline
  \ccpip Resonant & 86.0 & 9.4 \\ \hline
  \ccpip Coherent & 6.3 & 0.2 \\ \hline
  \ccqe & 2.4 & 85.4 \\ \hline
  Multi-pion & 2.5 & 0.02 \\ \hline
  \ccpin & 1.0 & 2.5 \\ \hline
  DIS & 0.2 & \(<\) 0.01 \\ \hline
  Other & 1.6 & 2.5 \\ \hline
\end{tabular}
\vspace{-2mm}
\caption{Predicted composition of \ccpiplike (one $\mu^-$ and one $\pi^+$ in the final state) and \ccqelike (one $\mu^-$ and no pions in the final state) events.}
\label{effective_comp}
\end{center}
\vspace{-6mm}
\end{table}

\vspace{0mm}To map reconstructed to true energy, we form a migration matrix \(A_{ij}\) representing the number of MC events in bin \(i\) of reconstructed energy and bin \(j\) of true energy.  We then normalize each reconstructed energy bin to unity to obtain an unsmearing matrix.  This is equivalent to a Bayesian approach discussed in \cite{D'Agostini}; it differs from the standard matrix inversion method in that the resulting unsmearing matrix is biased by the MC distribution used to generate it.  We account for this in our uncertainties by including a variation in the MC distribution used to generate the matrix.  Because we have good data/MC agreement, this effect is small.  The advantage of this method is that it avoids the problems of numerical instability and the magnification of statistical errors which occur in matrix inversion.  This unsmearing procedure also proved insensitive to variations in neutrino energy reconstruction, confirming that it performs as intended.

With all the correction terms put together, the cross section ratio in each energy bin $i$ is:
\vspace{-1.5mm}
\begin{equation}
\vspace{0mm}
\frac{\sigma_{1\pi^+,i}}{\sigma_{QE,i}} = \frac{\epsilon_{QE,i} * \sum_j{U_{1\pi^+,ij} * f_{1\pi^+,j} * N_{1\pi^+-cuts,j}}}{\epsilon_{1\pi^+,i} * \sum_j{U_{QE,ij} * f_{QE,j} * N_{QE-cuts,j}}}
\vspace{-2mm}
\end{equation}

\vspace{1mm}\noindent where subscript \(i\) runs over bins in true neutrino energy, subscript \(j\) indexes bins in reconstructed neutrino energy, $N_{X-cuts}$ denotes the number of events passing cuts for $X =$ CC$1\pi^+$, CCQE, \(f\) denotes a signal fraction, \(\epsilon\) denotes a cut efficiency, and \(U\) is a neutrino energy unsmearing matrix that acts on a reconstructed distribution to return the true distribution.

\begin{figure}[thb]
\vspace{-3mm}
\includegraphics[width=8.2cm, height=6.1cm]{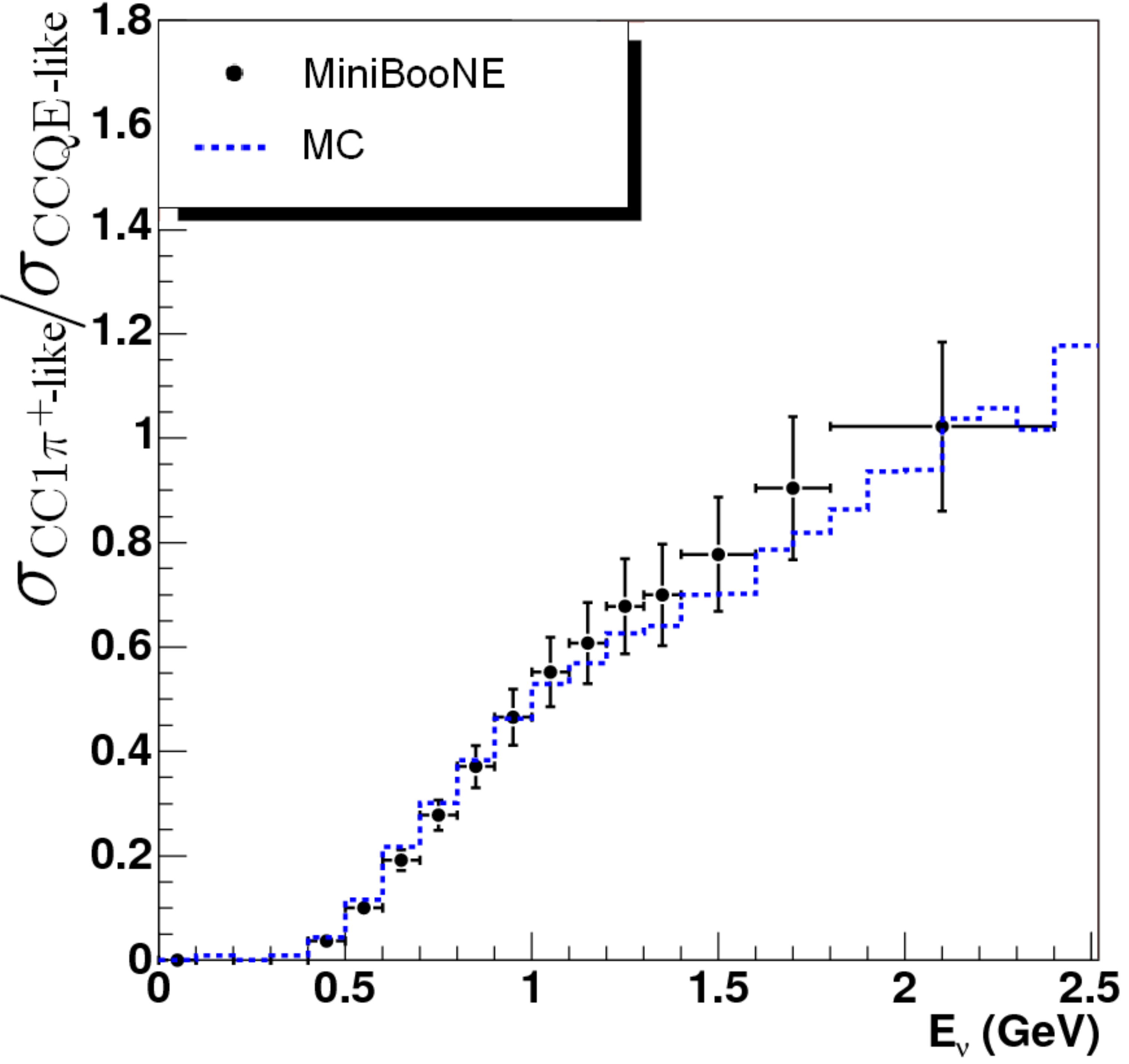}
\vspace{-4mm}
\caption{Observed CC$1\pi^+$-like/CCQE-like cross section ratio on $\mathrm{CH_2}$, including both statistical and systematic uncertainties, compared with the MC prediction \cite{NUANCE}.  The data have not been corrected for hadronic re-interactions.}
\label{fig:effective}
\vspace{-3mm}
\end{figure}

Figure \ref{fig:effective} shows the observed \ccpiplike to \ccqelike ratio extracted from the MiniBooNE data, including statistical and systematic uncertainties.

\vspace{-0.3mm}The dominant systematic uncertainties on the cross section ratio arise from four sources: the neutrino flux (which largely cancels in the ratio), the neutrino interaction cross sections (which affect the background predictions), hadron re-interactions in the detector, and the detector simulation (which describes light propagation in the oil). In the region of highest statistics (about 1 GeV), there is roughly an $8\%$ fractional error on the ratio resulting from hadron re-scattering in the detector, $6\%$ from neutrino cross sections, $4\%$ from the detector simulation, $2\%$ from the neutrino flux, and $2\%$ from the statistics of the two samples.

In addition to these errors, an uncertainty on the $Q^2$ dependence of the predicted \ccpip cross section is assessed based on comparison to MiniBooNE data. This contributes less than a 3\% overall error to the measured ratio. Additional variations testing the sensitivity of the result to the event selection scheme, reconstruction algorithm, energy unsmearing method, and predicted $\pi^+$ momentum distribution in \ccpip events are also included in the total uncertainty shown in Figure \ref{fig:effective}. Each of these added sources contributes a 1-2\% uncertainty to the ratio in the region of highest statistics.

\begin{figure}[htb]
\hspace{0cm}\includegraphics[width=8.2cm, height=6.1cm]{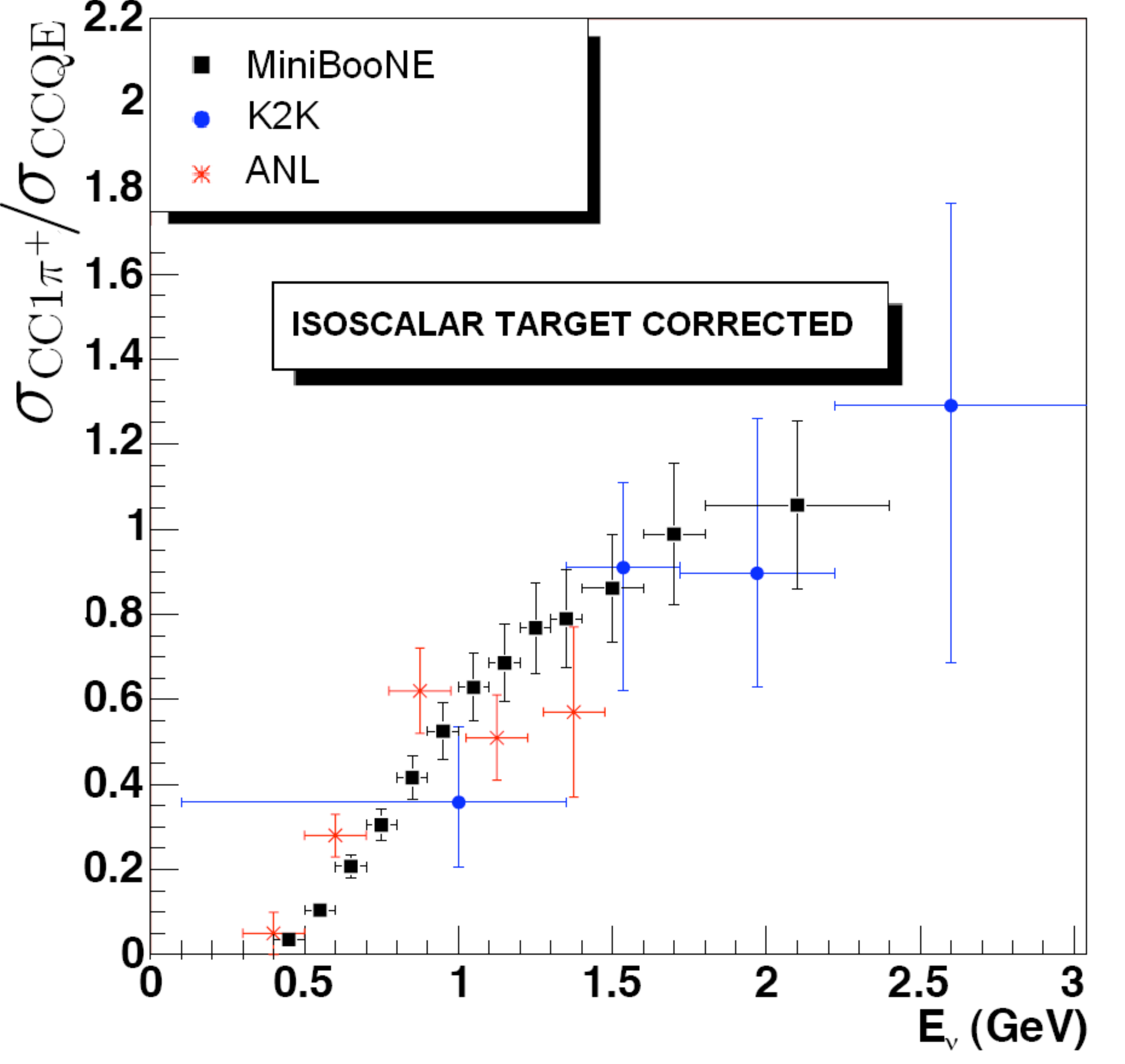}
\vspace{-4mm}
\caption{FSI-corrected \ccpip to \ccqe cross section ratio on $\mathrm{CH_2}$ compared with results from ANL ($D_2$) \cite{ANL} and K2K ($C_8H_8$) \cite{K2K}.  The data have been corrected for final state interactions and re-scaled for an isoscalar target.}
\label{fig:err_ratio_exp}
\vspace{-3mm}
\end{figure}

Unlike the result presented in Figure \ref{fig:effective}, the ratio reported by all prior experimental measurements \cite{ANL,BNL,K2K} has been one in which the effects of final state interactions (FSI) in the target nucleus have been removed using MC.  Solely for the purpose of comparison, we now extract a similarly corrected value. The FSI-corrected ratio is defined as the ratio of \ccpip to \ccqe events at the initial vertex and before any hadronic re-interactions. Thus, the signal fractions and cut efficiencies for the FSI-corrected ratio include corrections for intra-nuclear hadron re-scattering based on the MC's model for nuclear effects. The measurement proceeds exactly as for the observed ratio (Figure \ref{fig:effective}), except that now we define \ccpip and CCQE, rather than \ccpiplike and CCQE-like, events as signal for the respective samples. With these definitions, the CCQE (\ccpip) sample has a signal fraction of 72\% (87\%) and a cut efficiency of 37\% (20\%) in 500 cm. The FSI-corrected ratio is shown in Figure \ref{fig:err_ratio_exp}. The corrections for final state interactions have uncertainties associated with them, introducing additional systematic error to the cross section ratio.  The fractional error on the ratio due to these corrections is roughly 6\% in the region of highest statistics.

\vspace{-0mm}

Here we limit our comparison to those experiments which reported both CCQE and CC$1\pi^+$ cross sections, using the same energy bins for each of these interactions, so as to facilitate comparison with our measured CC$1\pi^+$/CCQE ratio.  Our result agrees with both ANL, which used a deuterium target, and K2K, which used $C_8 H_8$ (Fig. \ref{fig:err_ratio_exp}).  In order to make this comparison, the MiniBooNE and K2K results have been re-scaled to an isoscalar target. To perform this correction, we rescale the ratio by a factor of $(1-r)s_p$, where $r$ is the ratio of neutrons to protons in the target and $s_p$ is the fraction of $\pi^+$ production that is predicted (by MC) to occur on protons. The resulting scaling factor is 0.80 for MiniBooNE; for K2K we use the factor of 0.89 provided in \cite{K2K}. The results have not been corrected for their differing nuclear targets nor for the application of explicit invariant mass requirements (although the latter are similar). ANL used an explicit cut on invariant mass $W <$ 1.4 GeV. While no invariant mass cut is used in this analysis, the MiniBooNE spectrum is such that \ccpip events occur only in the region $W <$ 1.6 GeV; similarly, K2K's measurement covers the region $W <$ 2 GeV \cite{K2K}.

The dominant reason for the difference between the ratios presented in Figures \ref{fig:effective} and \ref{fig:err_ratio_exp} is intra-nuclear pion absorption in \ccpip events, which cause these events to look CCQE-like. As a result of $\pi^+$ absorption, a significant number of \ccpip events appearing in the numerator in Figure \ref{fig:err_ratio_exp} are in the denominator in Figure \ref{fig:effective}.  Thus, the FSI-corrected ratio, shown in Figure \ref{fig:err_ratio_exp}, is 15\% to 30\% higher than the observed ratio in our energy range.

\vspace{-0.2mm}

\vspace{-0.0mm}
In summary, MiniBooNE has measured the ratio of \ccpiplike to \ccqelike events for neutrinos with energy 0.4 GeV \(< E_\nu <\) 2.4 GeV incident on $\mathrm{CH_2}$. This is the first time such a ratio has been reported. Additionally, the ratio of the \ccpip and \ccqe cross sections at the initial vertex has been extracted using MC to remove the effects of final state interactions, in order to facilitate comparison with previous experimental measurements. The results are summarized in Table II. The measured ratios agree with prediction \cite{NUANCE,RS} and previous data \cite{ANL, K2K}.

\begin{table}[htb]
\begin{center}
\begin{tabular}{| c | c  | c  |}
	\hline
	$E_\nu$ \vspace{-1.2mm} & CC1$\pi^+$/CCQE &  CC1$\pi^+$-like/CCQE-like \\
	(GeV) & (FSI corrected)  & (observed)  \\ \hline \hline
	$0.45$ $\pm 0.05$ & $ 0.045$  $\pm 0.008$ & $0.036$  $\pm 0.005$ \\ \hline
	$0.55$ $\pm 0.05$ & $ 0.130$  $\pm 0.018$ & $0.100$  $\pm 0.011$ \\ \hline
	$0.65$ $\pm 0.05$ & $ 0.258$  $\pm 0.033$ & $0.191$  $\pm 0.019$ \\ \hline
	$0.75$ $\pm 0.05$ & $ 0.381$  $\pm 0.047$ & $0.278$  $\pm 0.028$ \\ \hline
	$0.85$ $\pm 0.05$ & $ 0.520$  $\pm 0.064$ & $0.371$  $\pm 0.040$ \\ \hline
	$0.95$ $\pm 0.05$ & $ 0.656$  $\pm 0.082$ & $0.465$  $\pm 0.053$ \\ \hline
	$1.05$ $\pm 0.05$ & $ 0.784$  $\pm 0.100$ & $0.551$  $\pm 0.066$ \\ \hline
	$1.15$ $\pm 0.05$ & $ 0.855$  $\pm 0.114$ & $0.607$  $\pm 0.077$ \\ \hline
	$1.25$ $\pm 0.05$ & $ 0.957$  $\pm 0.132$ & $0.677$  $\pm 0.091$ \\ \hline
	$1.35$ $\pm 0.05$ & $ 0.985$  $\pm 0.141$ & $0.700$  $\pm 0.097$ \\ \hline
	$1.5$ $\pm 0.1$  & $ 1.073$  $\pm 0.157$ & $0.777$  $\pm 0.109$ \\ \hline
	$1.7$ $\pm 0.1$  & $ 1.233$  $\pm 0.207$ & $0.904$  $\pm 0.137$ \\ \hline
	$2.1$ $\pm 0.3$  & $ 1.318$  $\pm 0.247$ & $1.022$  $\pm 0.161$ \\ \hline
\end{tabular}
\vspace{-2mm}
\caption{The MiniBooNE measured \ccpip to \ccqe (Figure \ref{fig:err_ratio_exp}) and \ccpiplike to \ccqelike (Figure \ref{fig:effective}) cross section ratios on $\mathrm{CH_2}$ including all sources of statistical and systematic uncertainty.}
\label{ratio_table}
\end{center}
\vspace{-7mm}
\end{table}

We wish to acknowledge the support of Fermilab, the National Science Foundation, and the Department of Energy in the construction, operation, and data analysis of the MiniBooNE experiment.

\end{document}